[Title Page]

# Exploring EFL Secondary Students' AI-generated Text Editing While Composition Writing

***David James Woo*** (Precious Blood Secondary School, Hong Kong, China)

***Yangyang Yu*** (Shanghai Jiao Tong University, Shanghai, China)

***Kai Guo*** (The University of Hong Kong, Hong Kong, China)

**Corresponding author**

- Name: Yangyang Yu
- Email address: florayu0209@sjtu.edu.cn
- Phone: +86 15202134271
- Postal address: School of Foreign Languages, Shanghai Jiao Tong University, No. 800 Dongchuan Road, Minhang District, Shanghai, China, 200240

**Declaration of conflicting interest**

The authors report there are no competing interests to declare.

**Data availability statement**

The data supporting this study's findings are available from the first author, David James Woo, upon reasonable request.

**Acknowledgements**

The authors express their gratitude to Dr. Chi Ho Yeung and Dr. Hengky Susanto for their indirect support of the study's data preparation and analysis.

# Exploring EFL Secondary Students' AI-generated Text Editing While Composition Writing

## Abstract

### Purpose
Generative artificial intelligence (AI) is transforming how English as a foreign language (EFL) students write. Still, little is known about how students manipulate text generated by generative AI during the writing process. This study investigates how EFL secondary school students integrate and modify AI-generated text when completing an expository writing task.

### Methodology
The study employed an exploratory mixed-methods design. Screen recordings were collected from 29 Hong Kong secondary school students who attended an AI-assisted writing workshop and recorded their screens while using generative AI to write an article. Content analysis with hierarchical coding and thematic analysis with a multiple case study approach were adopted to analyze the recordings.

### Findings
Fifteen types of AI-generated text edits across seven categories were identified from the recordings. Notably, AI-initiated edits from iOS and Google Docs emerged as unanticipated sources of AI-generated text. A thematic analysis revealed four patterns of students' editing behaviors based on planning and drafting direction: planning with top-down drafting and revising, top-down drafting and revising without planning, planning with bottom-up drafting and revising, and bottom-up drafting and revising without planning. Network graphs illustrate cases of each pattern, demonstrating that students' interactions with AI-generated text involve more complex cognitive processes than simple text insertion.

### Originality
The findings challenge assumptions about students' passive, simplistic use of generative AI tools and have implications for developing explicit instructional approaches to teaching AI-generated text editing strategies in EFL writing pedagogy.

## 1. Introduction

Generative artificial intelligence (AI) offers English as a foreign language (EFL) students access to not only an ideal partner for planning and revising written compositions (Su et al., 2023) but also a producer of high quality text that can be pasted to and manipulated in compositions. Although research has examined students' perceptions of writing with AI tools (Woo et al., 2024c) and the impact of AI-generated text on students' compositions (Woo et al., 2024b), we know little about how students actually manipulate AI-generated text when drafting a composition. This study investigates how EFL students in Hong Kong secondary schools edit AI-generated text when completing an expository writing task. By analyzing screen recordings, we identify the qualities of AI-generated text edits as well as editing patterns. The findings provide insight into how students engage in machine-in-the-loop writing (Clark et al., 2018), with unexpected findings about AI-initiated edits, and implications for developing appropriate expectations and effective pedagogical approaches for EFL writing instruction with AI-generated text.

## 2. Literature Review

### *2.1 Writing as a Cognitive Process*

Writing is a complex process that entails a combination of cognitive activities. Three key phases have been identified that serve as the foundation for writing research and pedagogy: *planning* for idea generation, organization or goal-setting, *drafting* for the translation of ideas into written language, and *reviewing* for improved text quality (Flower & Hayes, 1981). These activities are continuously mediated by a writer's internal resources such as memory, and the external task environment such as available tools, making writing a dynamic system open to individual and contextual influences (Hayes, 2012). Writers may thus vary in specific writing trajectories, embracing different writing experiences and demonstrating distinct writing performance. For instance, whereas experienced writers take time to create detailed outlines before drafting and engage in ongoing evaluation, novice writers often plan with minimal effort, if not skipping this phase, and refrain from making revisions (Becker, 2006).

For EFL learners who usually start as novice writers, the process of writing in a foreign language can be even more cognitively demanding. Not only do they encounter challenges shared with L1 writers in limited working memory and metacognitive regulation (Hayes, 2012), they also face greater “linguistic, rhetorical and ideational” difficulties because of language proficiency gaps and scant exposure to English writing (Manchón et al., 2009: 116). During the planning phase, lack of genre knowledge may prevent organizing key elements consistent with genre conventions (Han & Hiver, 2018). While drafting compositions, low-proficiency EFL writers tend to constantly stop to deliberate on grammar and vocabulary, thereby struggling with cognitive overload and leaving content underdeveloped (Kormos, 2012). Similarly, a pronounced review problem is that more attention is allocated to surface-level errors than global issues, which negatively impacts the final writing quality (Xu, 2018).

To scaffold EFL learners’ writing process, researchers have integrated various technological tools into writing training, such as digital mind-map for brainstorming (Karim et al., 2020), Google Docs for collaborative writing (Alsahil, 2024), and automated writing evaluation (AWE) for corrective feedback (Waer, 2023). Most tools primarily target specific cognitive activities in writing rather than influence the whole process. However, generative AI can be an exception, with the potential to bring comprehensive assistance across different writing stages (Su et al., 2023). It is worthwhile to explore how EFL learners write with generative AI and to what extent their writing process may be changed while interacting with this technology.

### *2.2. Machine-in-the-loop Writing*

‘Machine-in-the-loop’ writing refers to a collaborative composition process between a student and a generative AI interface (e.g., chatbot interface which simulates human turn-taking conversation). In this iterative process, the student first initiates with the AI interface. Based on the written instruction or prompt, the interface generates output which the student evaluates. The student makes deliberate decisions about the composition based on the output, including whether to integrate the output into the composition and whether to prompt the interface for writing support again. The student can also develop the composition without AI support. In this way, the student can plan, draft and revise a composition with a machine-in-the-loop while maintaining full agency over the writing process.

Studies have examined different phases of EFL students’ machine-in-the-loop writing, particularly through ChatGPT which has a chatbot interface. They have identified purposes for prompting (Woo et al., 2024a) and interventions that enhance students’ evaluation of chatbot output (Ngo & Hastie, 2025). Studies have also examined the impact of machine-in-the-loop writing on the planning, drafting and revision of compositions. Although studies show generative AI use can enhance EFL students' entire writing process (Boudouaia et al, 2024) and specific phases such as revision (Allen & Mizumoto, 2024), the *drafting* of compositions with AI-generated text has drawn the most concern from education stakeholders. For example, academics suggest EFL students may complete written tasks with little effort (Barrot, 2023), threatening academic integrity and creative thinking. Besides, EFL students fear losing their authorial voice (Wang, 2024). On the other hand, other studies have suggested that EFL students can be strategic users of AI-generated text when searching for ideas (Woo et al., 2023). Therefore, while generative AI is transforming the writing process of EFL students, we are only beginning to articulate students' actual behavior during the writing process, especially when drafting. This study aims to address the following research questions (RQs):

- ***RQ1***: What are the qualities of EFL secondary school students' edits to their compositions written with generative AI?
- ***RQ2***: What patterns emerge in how EFL students approach the editing of AI-generated text?

## 3. Methodology

### *3.1. Research Context*

In Hong Kong, students are typically taught expository and other genre writing through specific text types that may contain one or more genre features (Koh, 2015; The Curriculum Development Council, 2017). Furthermore, schools typically enroll students of similar academic proficiency (Lee & Chiu, 2017). To capture varied composition and editing qualities, we purposefully included seven Hong Kong secondary schools to compose an EFL student population that spans a wider range of academic achievement levels. Thus, we selected not only schools at the top-third of academic achievement but also at the bottom-third so that our findings can reflect student editing across Hong Kong secondary education's academic spectrum.

At each school, an English teacher was in charge of recruiting students and enrolling them in a free, two-hour workshop at their school. The teacher was also responsible for securing a venue and hardware for the workshop, as well as selecting the theoretical approach to EFL writing instruction, either a genre-based (Hyland, 2019) or a process-based (Flower &

Hayes, 1981) approach. The teacher-in-charge attended the workshop, observing it but not intervening in student performance.

The first author designed the instructional materials including a slide deck, worksheets, online quizzes and polling software. The author was also the instructor for the workshop. First, students reviewed knowledge on either process writing or genre writing. Second, students were introduced to prompt engineering knowledge, including types of natural language processing tasks (Ouyang et al., 2022) and prompt patterns (White et al., 2023). Third, the instructors and students collaborated to practice writing prompts for an expository writing task commonly found in the Hong Kong secondary school EFL writing curriculum. For a process-based approach, the instructor and students practiced writing prompts to plan an expository text's content, language and organization, to draft a section of the text, and to review the text. For a genre-based approach, the instructor and students practiced writing prompts to generate a model text and to analyze the model's content (see Figure 1), language and organization, to jointly construct the target text with AI by generating a template or a section of the text, and to independently construct the target text with AI generating writing plans and reviewing a completed text. The instructor and students practiced prompt engineering on the Platform for Open Exploration (POE) app (see Figure 2), which aggregates state-of-the-art generative AI chatbots (see Figure 3). POE is popularly used in Hong Kong because it provides free access to geo-blocked, American commercial chatbots such as ChatGPT, Claude and Gemini, which Hong Kong residents cannot directly access from OpenAI, Anthropic and Google websites respectively.

**Section 3 -- Guided Practice**

**Warmup: Setting the context**

Look at the picture below and read the writing task. Complete the statements below about the writing task's context.

Figure 7.

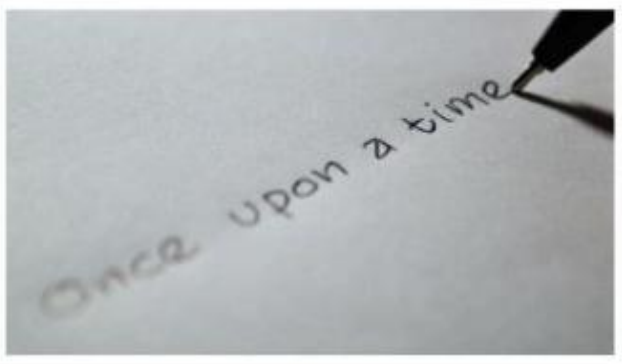

Contestants in a story-writing contest complained when a story created by artificial intelligence (AI) won first prize.

Write a letter to the editor on whether AI-generated stories should be allowed to compete in such contests. Write no more than 500 words.

The text type is a 1)________________________________.

The target reader or audience for this text type is 2)________________________________.

The purpose of this text is to 3)______________________________.

The style of this text is 4)________________________________.

The topic of this text is 5)________________________________.

What prompts could you write for ChatGPT's help to complete the above statements? Discuss with a partner.

In this section, you will learn to write a genre with ChatGPT support.

**Task: Modeling**

1. Work in pairs. Write a prompt for ChatGPT to *generate* a model text for the writing task. In addition, prompt ChatGPT to play a *persona*. If you are satisfied with ChatGPT's output, write your prompt below.

________________________________________________________________

________________________________________________________________

________________________________________________________________

**Figure 1.** A genre-based worksheet excerpt to practice prompt engineering

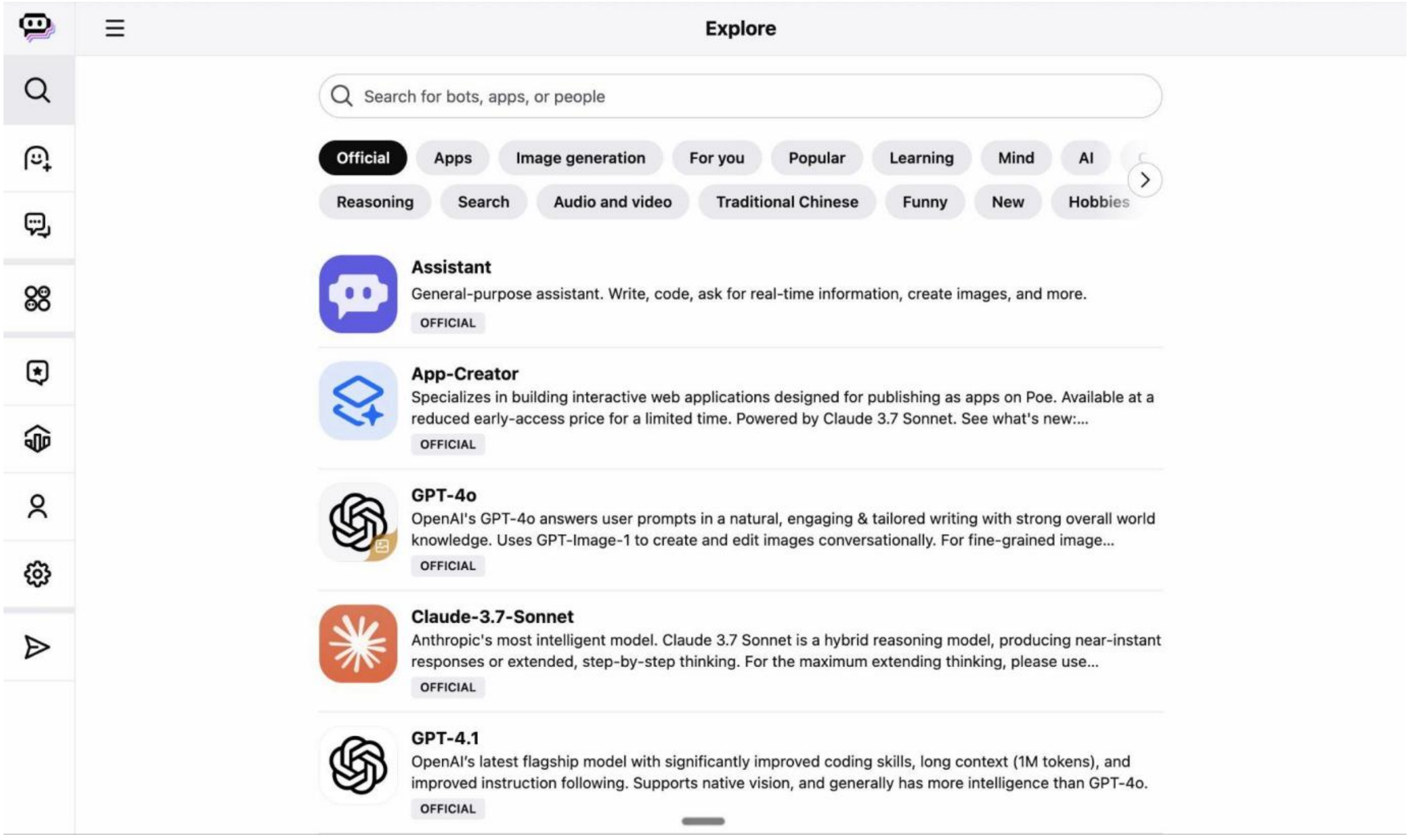

**Figure 2.** The POE app interface

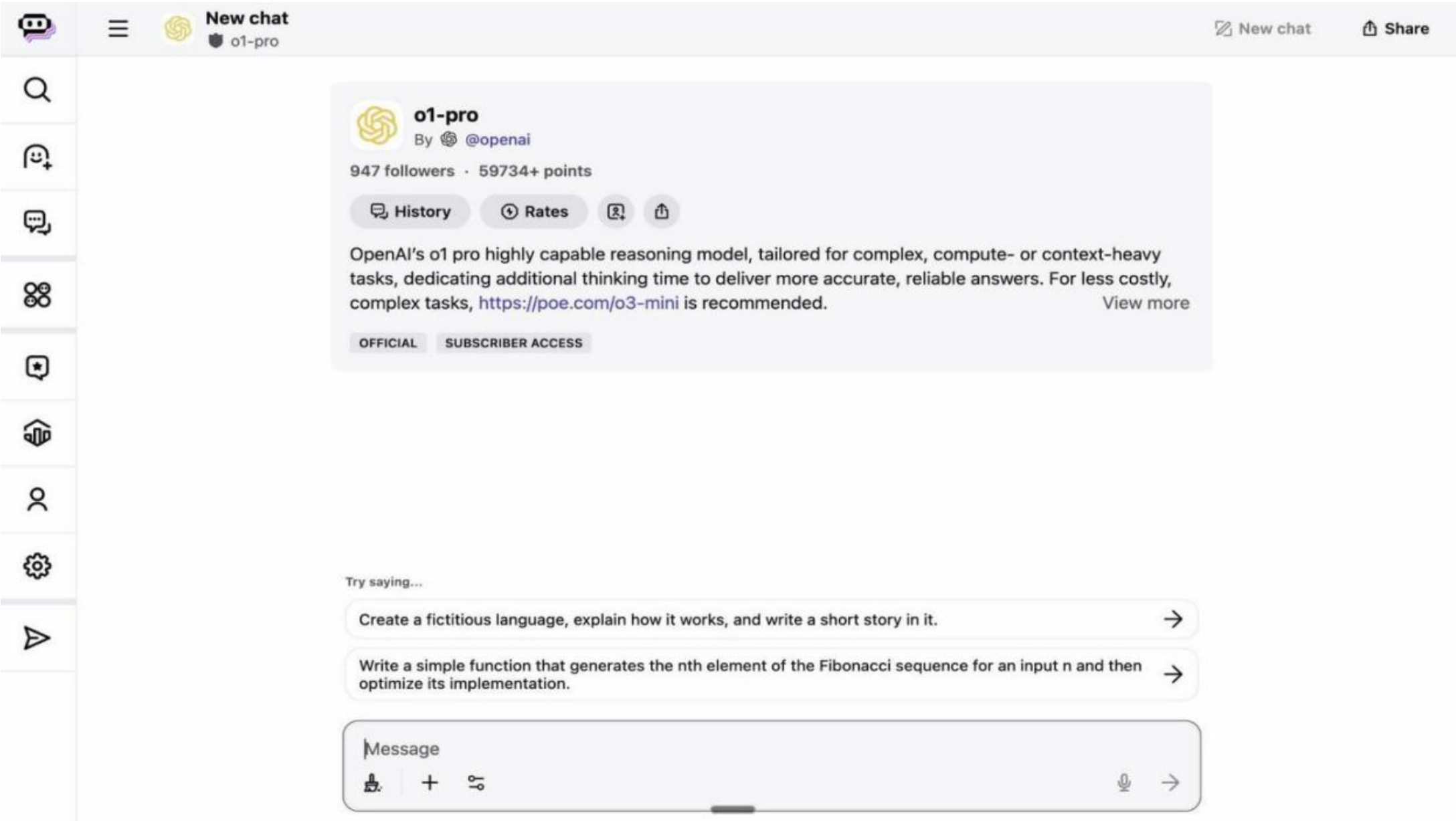

**Figure 3.** The interface for a state-of-the-art chatbot on the POE app

Lastly, students had at least 30 minutes to attempt an authentic, 500-word, article writing task taken from Hong Kong's secondary school exit examinations (see Figure 4). Students had to compose an article independent of classmate and teacher intervention, using any combination of self-written and AI-generated content from POE chatbots. Students wrote on Google Docs, highlighted AI-generated text in their compositions (see Figure 5). Students did not need to complete the task at the workshop.

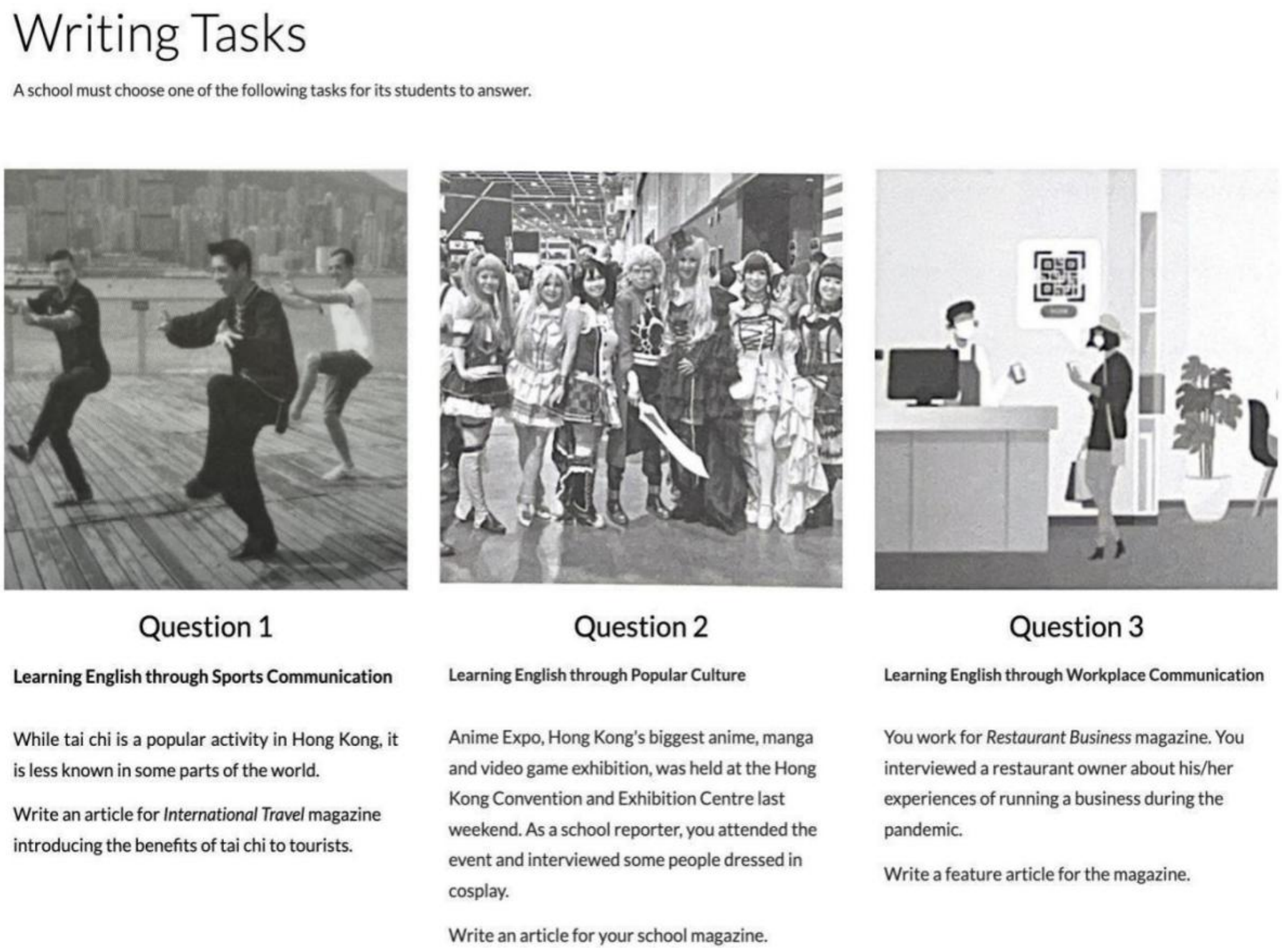

**Figure 4.** Three expository writing tasks

Title: Embracing Tranquillity: Discover the Timeless Art of Tai Chi in Hong Kong

In Hong Kong, Tai Chi holds a special place within the vibrant cityscape. Unlike other regions where it may be less phenomenal, Hong Kong provides an unparalleled opportunity to immerse oneself in the authentic practice of Tai Chi and witness its influence on the local community. Each morning, the city awakens to the rhythmic movements of Tai Chi practitioners from the waterfront promenades to the lush parks.

Tai Chi's popularity in Hong Kong stems from its ability to provide a wide range of benefits, encompassing both physical and mental well-being. Its slow and controlled movements improve flexibility, balance, and cardiovascular health while reducing chances of getting injuries. Regular practice also fosters mental clarity, lowers stress, and cultivates inner peace. In addition, Tai Chi holds deep cultural significance as it originated from ancient Chinese martial arts traditions, embodying principles of balance, harmony and cultivating life energy known as "Qi." You will have a unique opportunity to connect with Chinese culture on a deeper level through Tai Chi.

Moreover, Tai Chi sessions allow you to connect with the locals' daily lives, fostering new friendships and cultural exchanges. The shared experience of immersing oneself in the gentle movements and peaceful atmosphere creates lasting memories and a sense of belonging. You now have the wonderful opportunity to participate in daily Tai Chi sessions benefiting both your physical well-being and mental clarity.

It's worth mentioning that celebrities like action star Jackie Chan, Hollywood icon Robert Downey Jr., and actress Helen Mirren have incorporated Tai Chi into their wellness routines, recognizing its transformative power. Their dedication to this ancient art form not only attests to its growing popularity but also serves as an inspiration for travellers seeking to explore its benefits.

**Figure 5.** An article excerpt for Question 1 with AI-generated text highlighted in red

### *3.2. Participants*

The researchers informed students of the study and their rights. Twenty-nine students voluntarily agreed to participate. The students ranged in age from 12 to 18, coming from secondary school grade level one to grade level five.

### *3.3. Data Collection*

To collect data on EFL students' drafting and revision with AI-generated text, students recorded their screens while completing the expository writing task during the workshop. The researchers collected the screen recordings before students left the workshop.

### *3.4. Data Analysis*

We employed a mixed-methods analytical approach. Table 1 shows the alignment of RQs with analytical approaches and methods. The following sub-sections elaborate each analytical approach.

**Table 1**. RQs aligned with data analysis framework

| RQ | Analytical approach | Methods |
|---|---|---|
| RQ1 | Content analysis with hierarchical coding | Provisional coding for expository writing units (title, heading, topic sentences, etc.)<br>Coding for syntactic complexity (short, medium, long)<br>In situ coding for edit types<br>Axial coding to identify categories<br>Descriptive statistics of edit frequencies |
| RQ2 | Thematic analysis with a multiple case study approach | Review of observation notes and memos<br>Development of process codes<br>Integration of codes into themes<br>Case study analysis with theoretical replication<br>Network graph visualization |

*3.4.1. Content analysis with hierarchical coding*

To answer RQ1, we performed a directed and conventional content analysis (Hsieh & Shannon, 2005), developing a hierarchical coding scheme comprising three dimensions: expository writing units; syntactic complexity; and edit types. The development of codes for expository writing units and syntactic complexity was directed. Codes for expository writing features were based on Hong Kong secondary school EFL textbooks (Potter et al., 2023; Nancarrow & Armstrong, 2024): title; introductory paragraph; heading; body paragraph topic sentence; body paragraph supporting sentence; and concluding paragraph. Three codes for syntactic complexity (Lu, 2010) were based on production unit lengths: short (less than one unit); medium (exactly one unit); and long (exceeding one unit). The development of codes for edit types was conventional, made in situ (Saldana, 2012), by viewing screen recordings, following an observational protocol and making notes to identify and describe each edit's insertion, deletion or modification. We iteratively consolidated these emerging codes and their descriptions.

During the initial coding phase, for each student's screen recording, we first noted the runtime, and the student's school, form level, and unique identifier. We observed each edit in a screen recording, noting the time of the edit and its sequence. We simultaneously coded the edit for an expository unit, syntactic length and type of insertion, deletion or modification. Per our observation protocol, we noted what actions preceded and followed each edit.

We employed axial coding to identify relationships between open codes, consolidating them into broader categories. To ensure reliability, the first author coded all recordings while the second author independently coded 15% of the recordings, achieving 100% agreement through discussion and codebook refinement.

### *3.4.2. Thematic analysis with multiple case study*

To answer RQ2, we conducted thematic analysis (Braun & Clarke, 2006) to identify patterns in students' editing approaches. During the initial coding phase, we wrote reflective memos with initial thoughts on emerging codes for types of AI-generated text insertion, deletion or modification and possible themes from the observation notes. Reviewing reflective memos and observation notes, we developed process codes that we integrated into themes. We present these themes using a multiple case study approach (Yin, 2009), treating each student's editing process as a descriptive case with theoretical replicability.

We visualized cases through network graphs (Miles et al., 2013). A graph illustrates relationships among a student's AI-generated text edits with consideration to the type, sequence and syntactic complexity of the edits. Nodes represent edit types from the axial coding, edges show directionality between edits, and numbers indicate sequential order. The number's color indicates the edit's syntactic complexity. These graphs provided both within-case and cross-case displays, highlighting pattern differences and variations.

## 4. Results

Twenty-five out of 29 participants' screen recordings showed AI-generated text edits and were analyzed. The total runtime of the 25 screen recordings was 10 hours, 34 seconds and the average length was 24 minutes. The longest was 35 minutes, 22 seconds and the shortest 1 minute, 6 seconds.

### *4.1. What are the qualities of EFL secondary school students' edits to their compositions written with generative AI? (RQ1)*

Table 2 presents 15 codes that specify types of AI-generated text insertions, deletions and modifications from the screen recordings. We organized these 15 codes into seven categories (number of codes): 1) insertions of AI-generated text (n=3) that did not displace other text in a composition; 2) deletions of AI-generated text (n=1) without replacing that text with other text; 3) AI-generated text replacements (n=3) where students replaced AI-generated text in their composition with either output from a chatbot, a student's own words or text from other sources such as the writing task prompt; 4) human text replacements (n=3) where students replaced either their own words or a combination of their own words and AI-generated text with either chatbot output or existing AI-generated text in their composition; 5) pre-edit and formats (n=1) where students highlighted AI-generated text in a composition for subsequent editing or other purposes; 6) Apple iOS predictive text edits (n=1) where on some iOS keyboards students could click on next word predictions and iOS would automatically perform the insertion of those suggested words; and 7) Google Docs edits (n=3) where for some students Google Docs would automatically underline words and suggest correct spelling or grammar for those words or show next word predictions. Google Docs would automatically perform the edit on text if a student clicked on the suggestion for spelling and grammar edits, or hit tab for next word prediction. Apple iOS and Google Docs were unanticipated sources of AI-generated output for compositions.

We coded 266 edits and Table 2 provides a frequency count and student distribution by types of AI-generated text insertions, deletions and modifications. Insertion (code no. 2) was the most frequently applied (n=68) edit type with the largest number of students (n=22). Figure 6 presents two screenshots with the top screenshot illustrating before an insertion edit and the bottom screenshot illustrating after. The second most frequently applied edit type was Deletion (code no. 4) (n=58) with the second largest number of students (n=13). Figure 7 presents two screenshots with the top screenshot illustrating before a deletion edit and the bottom screenshot illustrating after the deletion of the "Introduction" heading in the text. The third most frequently applied edit type was Replacement AI:Human (code no. 7) (n=44) with the third largest (n=12). Figure 8 presents two screenshots with the top screenshot illustrating before a replacement AI:human edit with the AI-generated text "The Event" highlighted, and the bottom screenshot illustrating after the replacement of the highlighted text with "It" in black color. The least frequently applied edit types were Replacement AI:Other (code no. 5) (n=1) with the smallest number of students (n=1), followed by Replacement Human-AI:AI output (code no. 10) (n=2) with the second smallest number of students (n=2) and Google: Spell check (code no. 15) (n=2) with the smallest number (n=1).

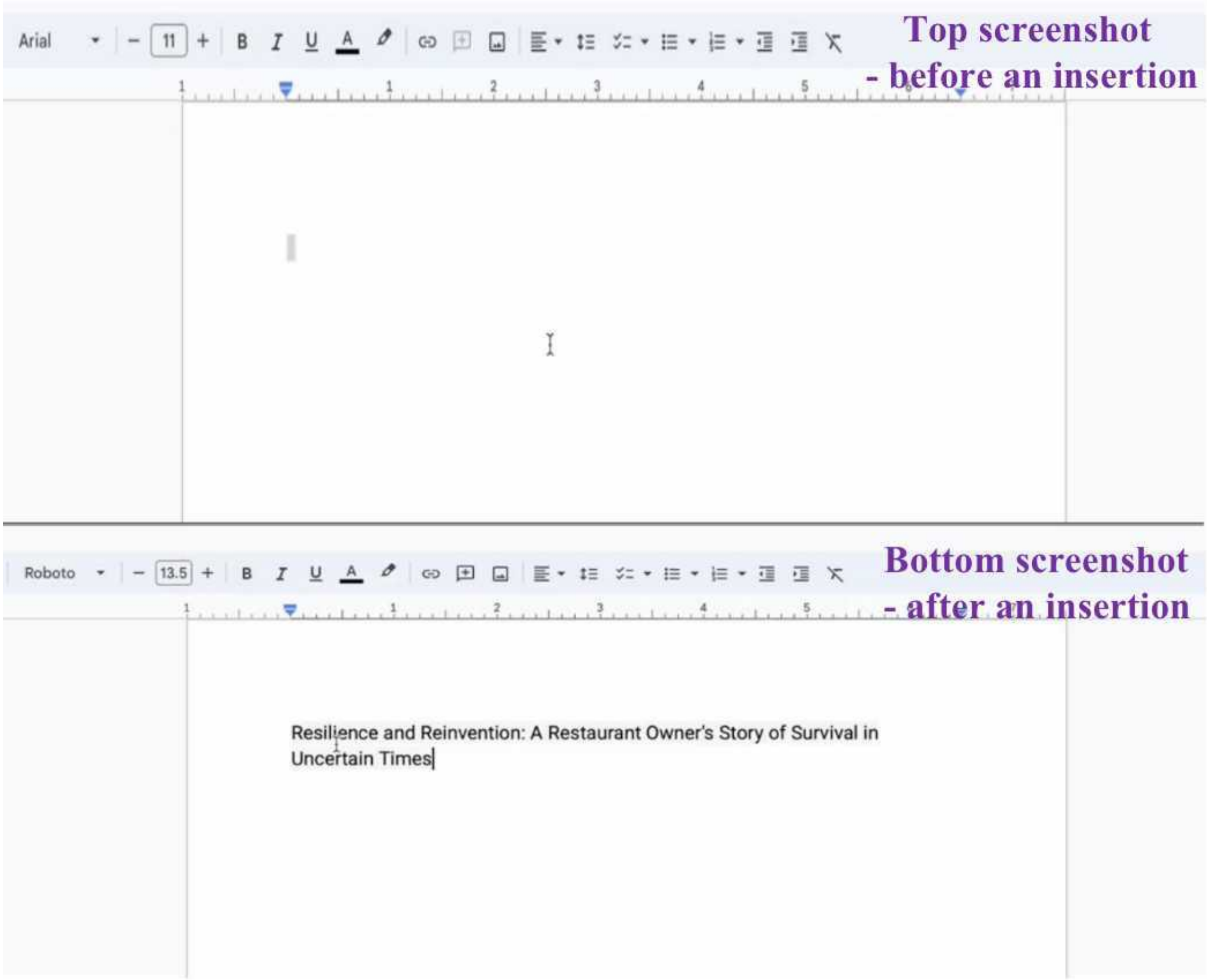

**Figure 6.** Screenshots of before and after an insertion edit

**Table 2.** 15 Types of AI-generated text insertions, deletions and modifications

| Category No. | Category (Abbreviation) and Description | Code No. | Code (Abbreviation) | Description | Frequency | No. of Students |
|---|---|---|---|---|---|---|
| 1 | Insertions (Is) - Inserting onto Google Docs AI-generated output from a chatbot, AI-generated text already on Google Docs or text from another source. | 1 | Cut and Paste (CP) | Removing AI-generated text and inserting it elsewhere. | 6 | 2 |
| | | 2 | Insertion (I) | Adding AI-generated output without replacement. This includes inserting complete AI-generated articles onto Google Docs. | 68 | 22 |
| | | 3 | Insertion: Other (IO) | Inserting copied text that is neither AI-generated text nor output, for instance, from the writing task prompt. | 4 | 2 |
| 2 | Deletion (D) - Removing AI-generated text without replacement. | 4 | Deletion (D) | Removing AI-generated text without replacement. This can include deleting a letter from an AI-generated word to appropriately conjugate a verb, for example. | 58 | 13 |
| 3 | Replacement AI (RA) - Replacing AI-generated text on the Google Doc with human-generated text, AI-generated output from chatbot or text from other sources; no replacement of AI-generated text with other AI-generated text was observed. Transforming an AI-generated semantic unit (e.g. heading) into another semantic unit (e.g. topic sentence) can be coded here. | 5 | Replacement AI:Other (RAO) | Swapping AI-generated text with copied text that is neither AI-generated text nor output, for instance, from the writing task prompt | 1 | 1 |
| | | 6 | Replacement AI: AI output (RAAO) | Swapping AI-generated text with AI-generated output | 8 | 6 |
| | | 7 | Replacement AI:Human (RAH) | Swapping AI-generated text with human text. This can include deleting an AI-generated word and rewriting the word with an -s suffix to correct verb conjugation, for example. Or replacing letters in a word. | 44 | 12 |
| 4 | Replacement Human (RH) - | 8 | Replacement Human:AI (RHA) | Swapping human text with AI-generated text. | 9 | 5 |
| | Replacing human-generated text on Google Docs with AI-generated output from chatbots or AI-generated text on Google Docs. | 9 | Replacement Human:AI output (RHAO) | Swapping human text with AI-generated output. | 3 | 3 |
| | | 10 | Replacement Human-AI:AI output (RHAAO) | Swapping human text and AI-generated text with AI-generated output or text. | 2 | 2 |
| 5 | Pre-edit and format (P/F) - Highlighting AI-generated text in a composition for subsequent editing or other purposes. | 11 | Format (F)<br>Pre-edit (P) | Highlighting (e.g. coloring, italicizing or making bold-face) AI-generated text for subsequent deletion, insertion or other purposes unrelated to spacing. | 7 | 2 |
| 6 | Apple (A) -Accepting any Apple iOS predictive text suggestion, such as that shown when typing on an iPad keyboard. | 12 | Apple (A) | Accepting any Apple iOS predictive text suggestion shown on an Apple digital keyboard on iPad. | 30 | 4 |
| 7 | Google (G) - Accepting any Google Docs suggestion, be it for correcting grammar or spelling or for predicting the next word(s). | 13 | Google: Grammar check (GG) | Accepting any Google Docs suggestion for grammar correction. | 8 | 4 |
| | | 14 | Google: Insertion (GI) | Accepting any Google Docs suggestion for next word prediction. | 16 | 3 |
| | | 15 | Google: Spell check (GS) | Accepting any Google Docs suggestion for spelling correction. | 2 | 1 |

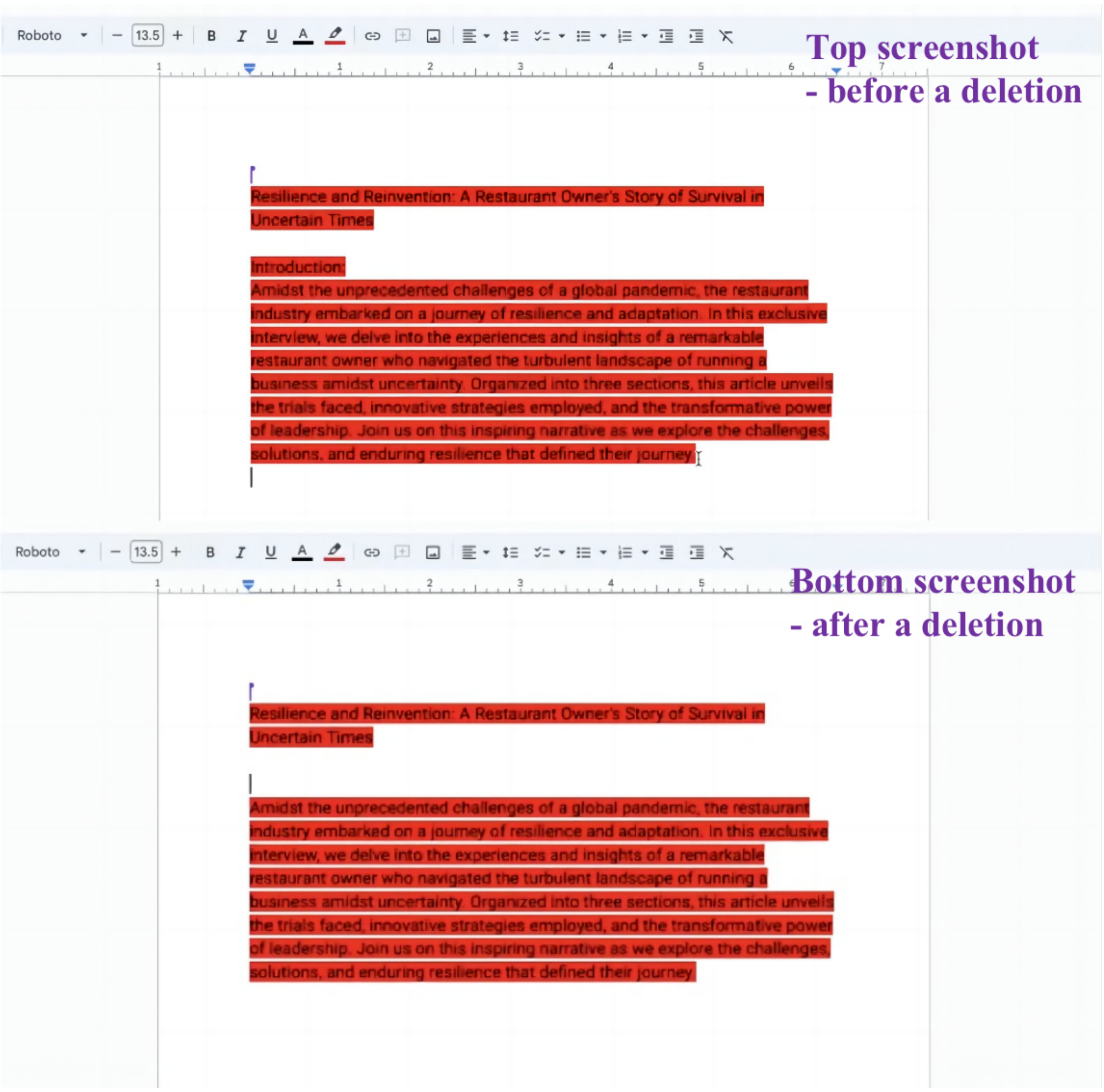

**Figure 7.** Screenshots of before and after a deletion edit

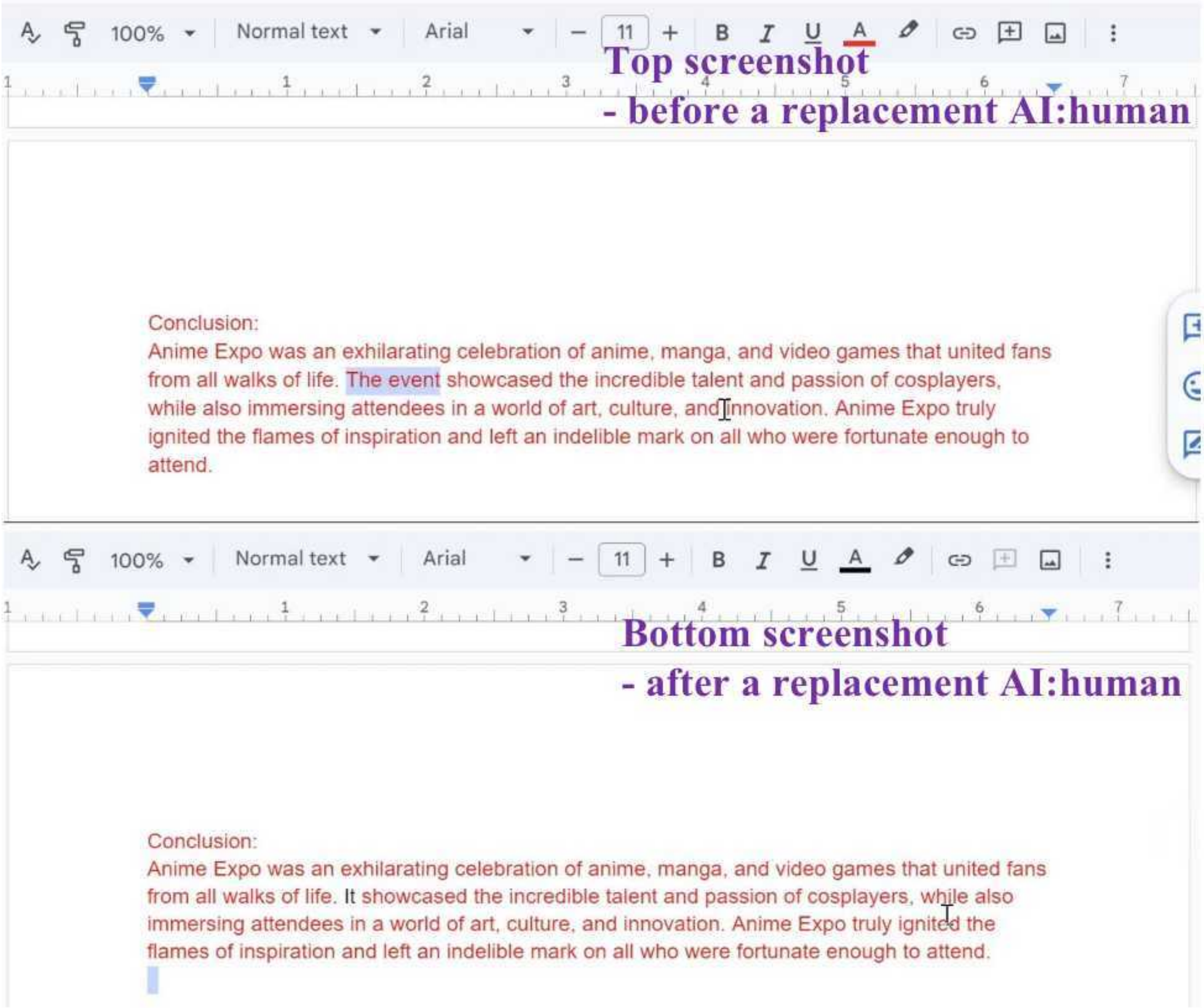

**Figure 8.** Screenshots of before and after a replacement AI:human edit

### *4.2. What patterns emerge in how EFL students approach the editing of AI-generated text? (RQ2)*

From our thematic analysis of observation notes and reflective memos, we developed two sets of binary process codes: *planning* or *no planning*; and *top-down drafting and revising* or *bottom-up drafting and revising*. Table 3 describes these drafting and revising; and planning codes. Figure 6 illustrates the bottom-up drafting and revising code as the student did not paste an article in an initial turn. Figure 9 presents two screenshots that illustrate the top-down drafting and revising code as the top screenshot shows a student highlighting their own words in an initial turn and the bottom screenshot shows those words replaced with an AI-generated article. Figure 10 illustrates the planning code as a student prompted ChatGPT on the POE app for a tree diagram.

**Table 3.** Drafting and planning codes

| Category | Code | Description |
|---|---|---|
| Drafting and Revising | Bottom-up drafting and revising | Drafting an article without pasting article paragraphs in an initial turn. |
| | Top-down drafting and revising | Pasting article paragraphs (i.e. introductory paragraph, body paragraphs and concluding paragraph), with or without title and headings, in an initial turn; pasting an article template is also a top-down approach and does not exhibit planning. |
| Planning | Planning | Any inquiry into the content, language or organization for the writing task before drafting; this includes referring to pre-existing notes. |
| | No planning | Absence of inquiry into the content, language or organization for the writing task before drafting. |

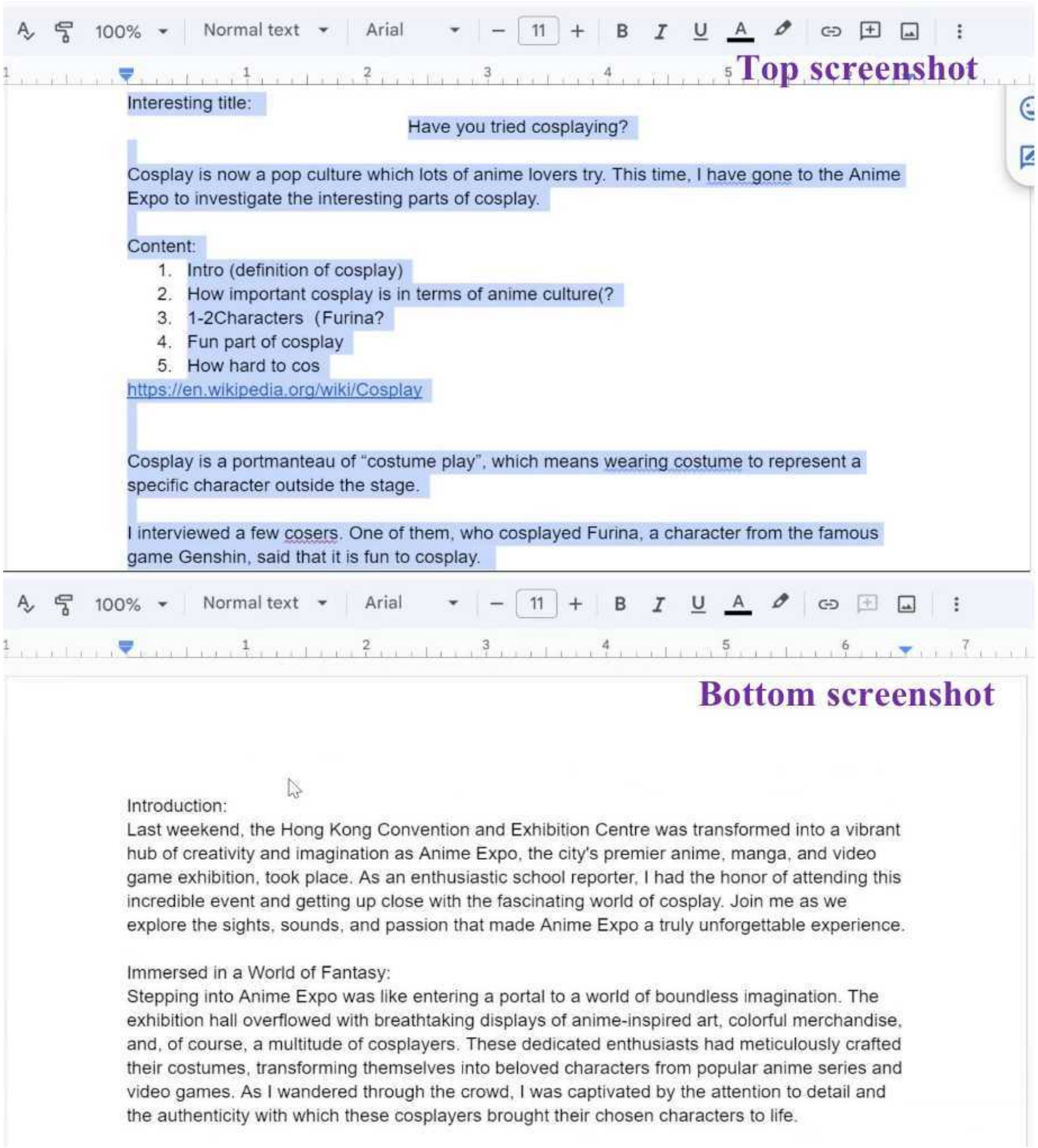

**Figure 9.** Screenshots illustrating top-down drafting and revising

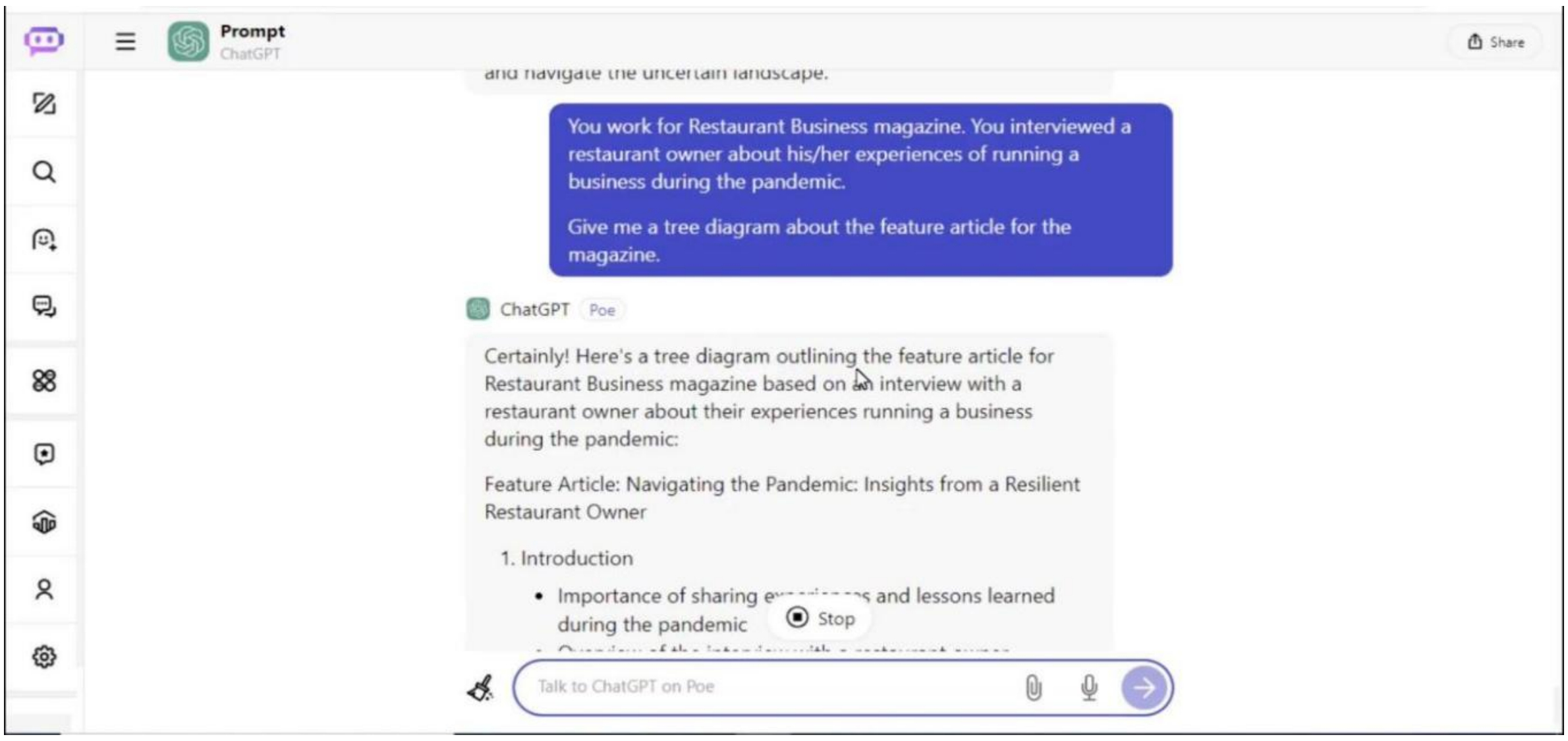

**Figure 10.** A screenshot illustrating planning

Integrating *planning* or *no planning*, and *top-down drafting and revising* or *bottom-up drafting and revising* codes, we arrived at four themes that can holistically describe the quality of students' editing patterns of AI-generated text. Table 4 shows the student distribution for these themes. The largest number of students was found planning with bottom-up drafting and revising and the smallest number found planning with top-down drafting and revising. We elaborate the themes with cases and network graphs.

**Table 4.** Student distribution for four themes

| | Planning | No Planning |
|---|---|---|
| Top-down drafting and revising | 1 | 6 |
| Bottom-up drafting and revising | 15 | 3 |

*4.2.1. Planning with top-down drafting and revising (n=1)*

Figure 11 illustrates the case of School II student 4D01's editing. The case features *planning*, inquiry into the writing task's content, language or organization before drafting. In this case, the student prompted ChatGPT for difficulties of running a restaurant during the pandemic, and then, based on those difficulties, a writing plan, which the student pasted to the composition in turns 1 and 2. The case also features *a top-down approach to drafting and revising*: the student pasted a complete set of article paragraphs (i.e., introductory paragraph, body paragraphs and concluding paragraph) and headings in an initial turn (turn 3); and then edited that AI-generated article. Specifically, in Figure 11, the student made short chunk deletions and replacements from the introductory paragraph (turns 4-6), to the topic sentence (turn 7) and the supporting sentences (turns 8-10) of the first body paragraph, and next to the topic (turn 11) and supporting sentences (turns 12-15) of the second body paragraph and the same to the third body paragraph (turns 16-19). After editing the concluding paragraph (turns 20-22), the student made subsequent edits, for instance, by deleting the writing plan (turns 23-25) and parts of headings (turns 26-31) from the initial pasting.

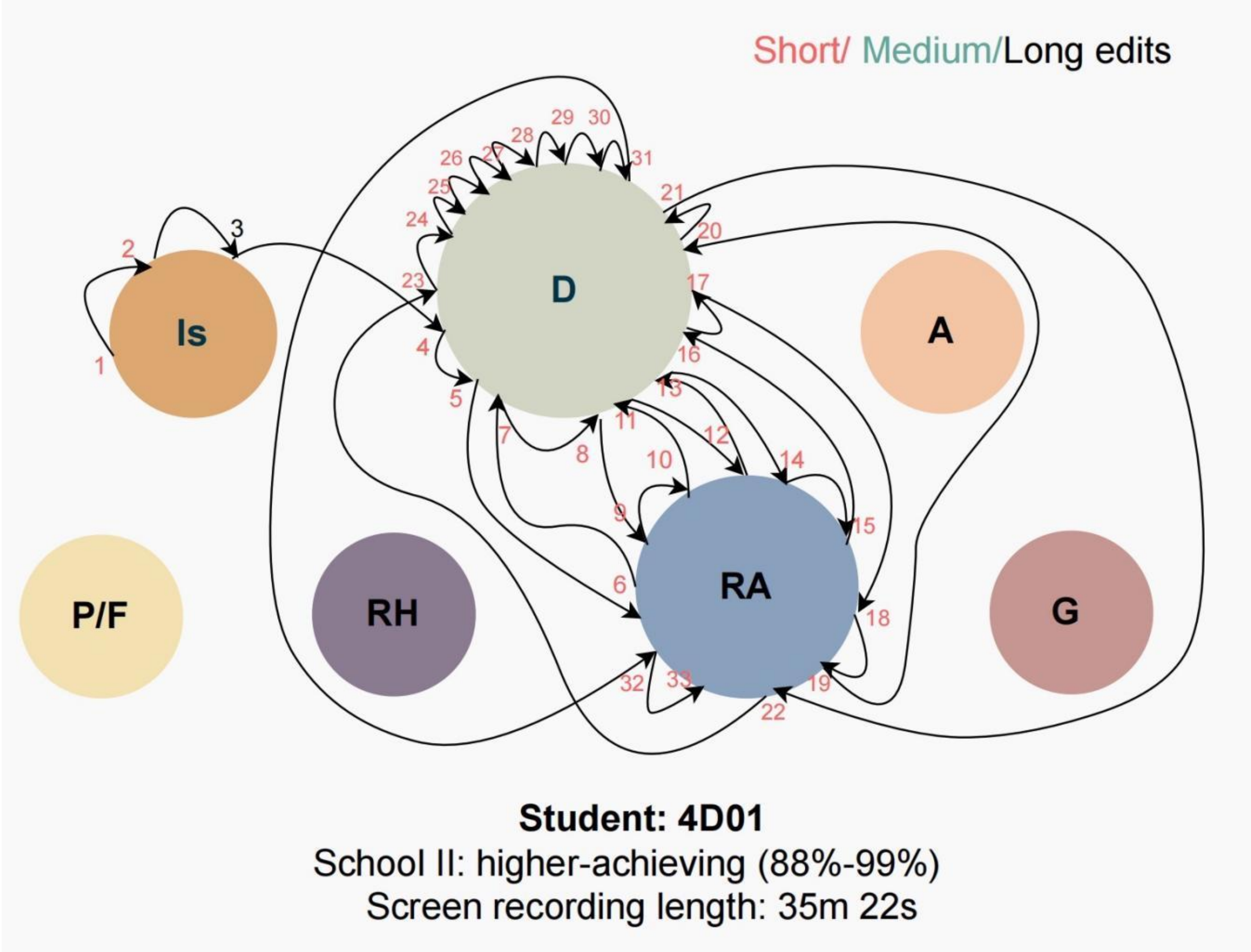

**Figure 11.** A case of planning with top-down drafting and revising

*4.2.2. Top-down drafting and revising without planning (n=6)*

Figure 12 illustrates the case of School VII student 4B23's editing. Like the case in Figure 11, this case features *a top-down approach to drafting and revising*. However, this case shows *no planning* as the student pasted a complete set of article paragraphs and headings from the start (turn 1) and began making edits from the introductory paragraph's heading (turn 2) to the title (turns 3-4) and to the introductory paragraph (turns 5-6) onwards. Compared to the case in Figure 11, this student made fewer total edits in the screen recordings and edited short, medium and long chunks.

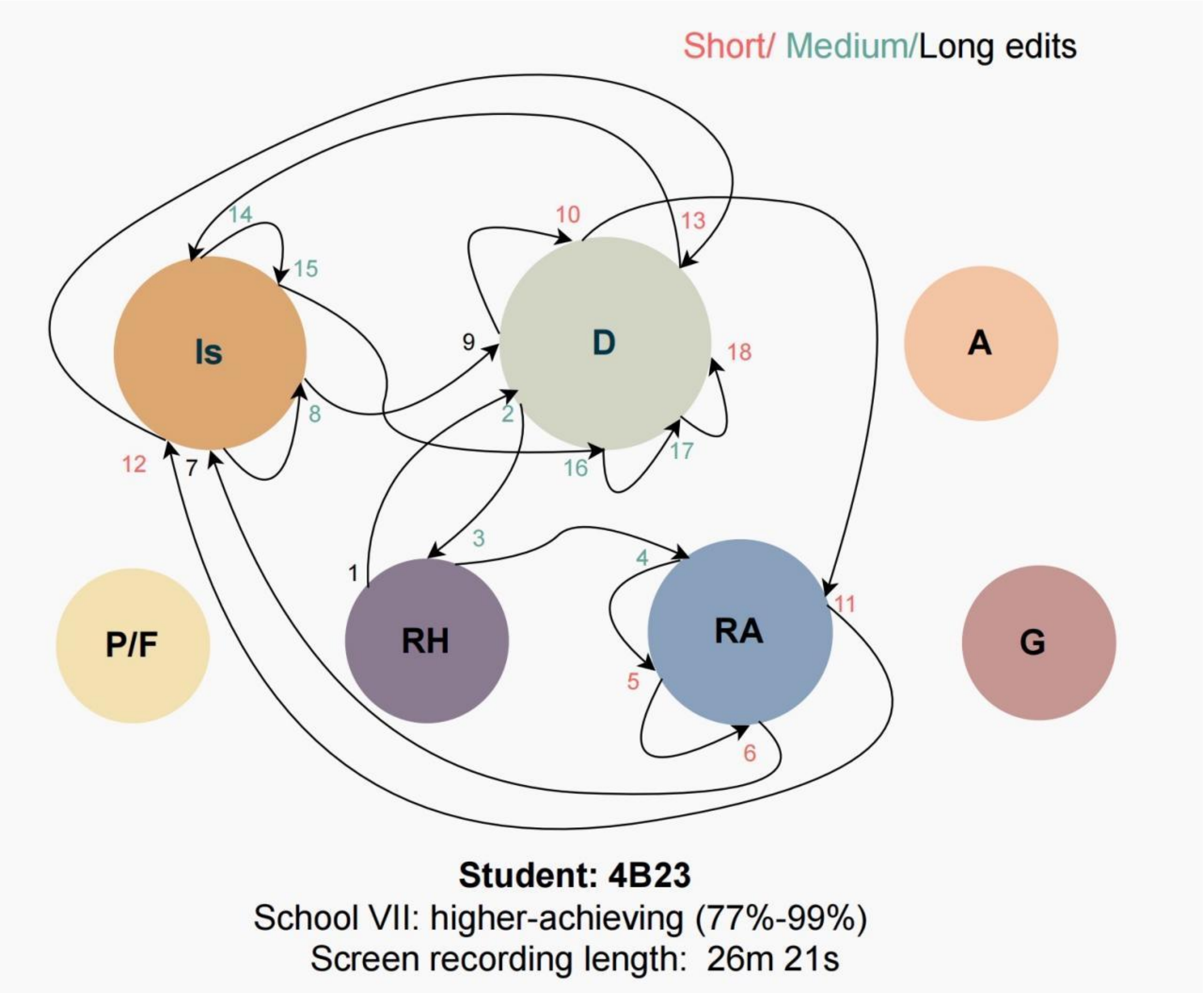

**Figure 12.** A case of top-down drafting and revising without planning

*4.2.3. Planning with bottom-up drafting and revising (n=13)*

Figure 13 illustrates the case of School II student 4D02's editing. The case features *planning* in that before drafting, the student asked ChatGPT how the pandemic affected the restaurant industry. The case features *a bottom-up approach to drafting and revising*. Using this approach, a student did not compose an article by pasting a complete set of article paragraphs in an initial turn. Instead, the student may develop the article with her own words, or by pasting an AI-generated expository writing unit and subsequently editing that unit. For the case in Figure 13, the student pasted an introductory paragraph to the composition (turn 1) and then began writing the topic sentence of the first body paragraph (turns 2-6) with her own words, Apple iOS predictive text suggestions and Google Docs suggestions. The student then wrote the supporting sentences of the first body paragraph in her own words and with Apple and Google AI-generated text (turns 7-14) etc. The student inserted two additional AI-generated body paragraphs (turn 15) and a concluding paragraph (turn 17).

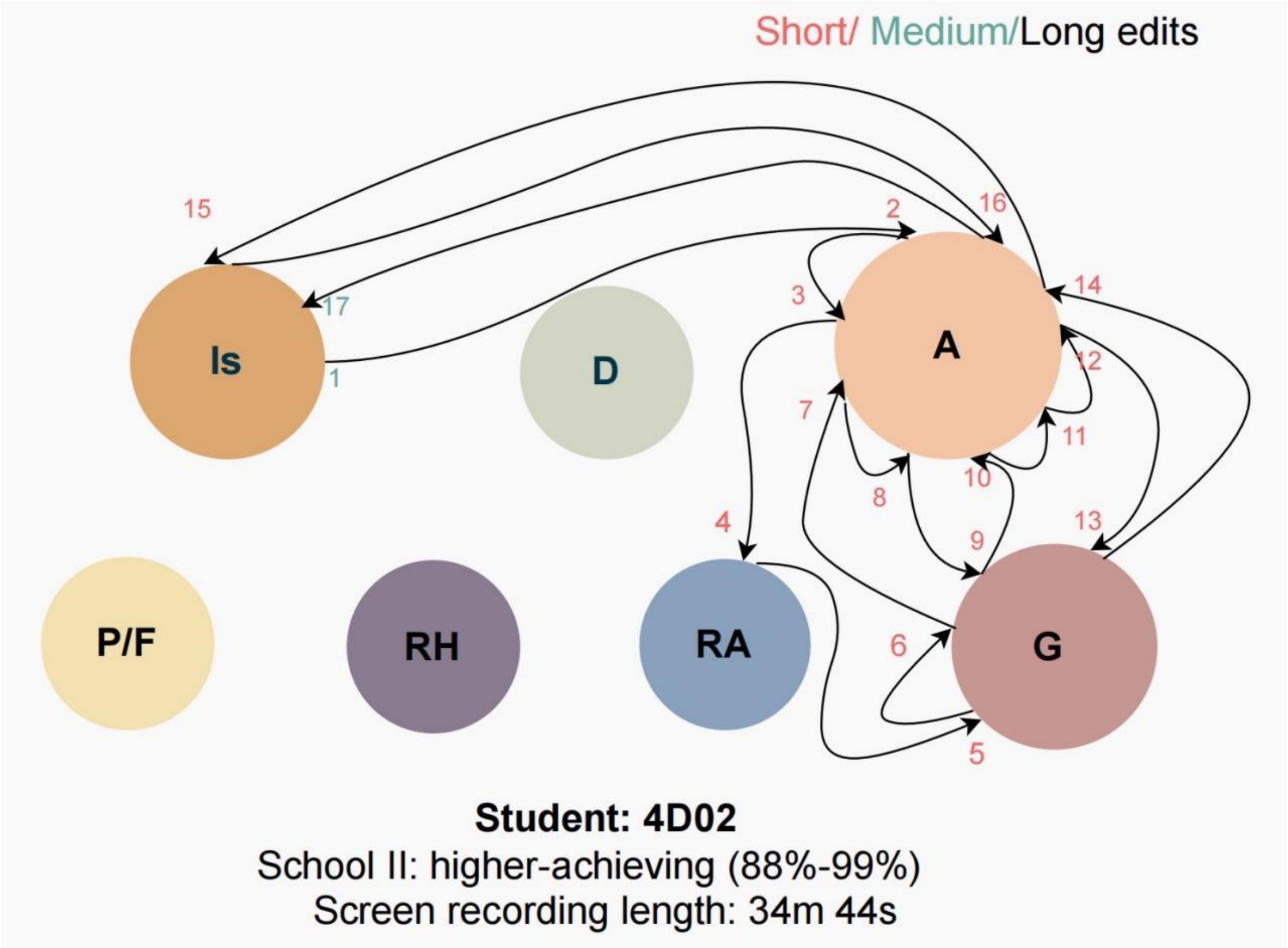

**Figure 13.** A case of planning with bottom-up drafting and revising

Figure 14 illustrates the case of School III student 5C17's editing. The case features the same *planning* with *bottom-up drafting and revising* pattern as illustrated in Figure 13. However, in the planning phase, this student prompted ChatGPT for an article example, instruction on how to write a heading, and suggestions for a title based on the writing task prompt. Furthermore, the student's screen recording shows only three turns, when the student pasted one of the titles suggested by ChatGPT (turn 1), pasted an introductory paragraph with a heading (turn 2) and then deleted that heading (turn 3).

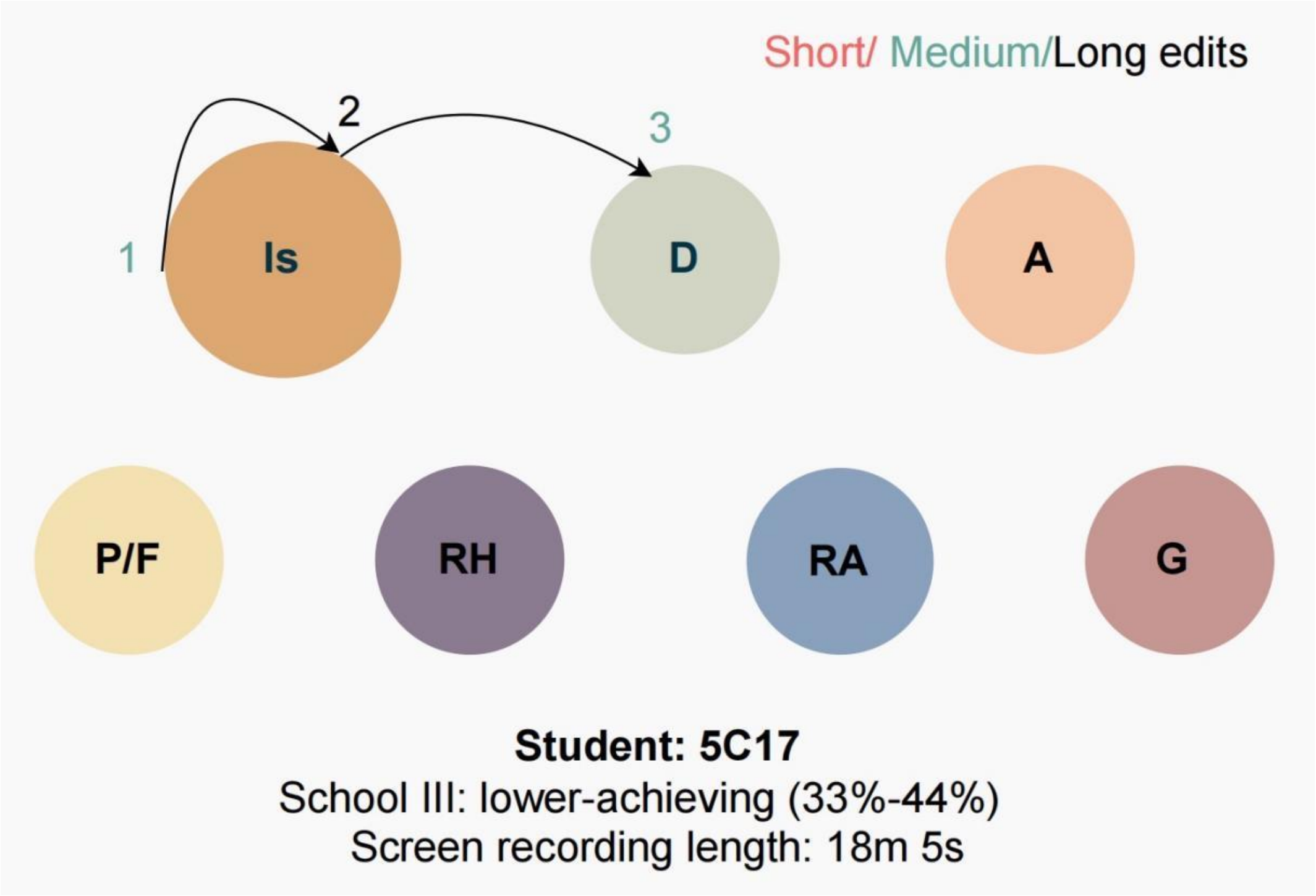

**Figure 14.** Another case of planning with bottom-up drafting and revising

Figure 15 illustrates the case of School III student 5B20's editing. The case features *planning* with *bottom-up drafting and revising*. To plan, the student prompted ChatGPT for hardships and challenges faced by restaurant owners during the pandemic, pasting that output comprising 10 bullet points and additional paragraphs to the composition (turn 1). The student then prompted another chatbot for hardships and challenges and for strategies that restaurant owners used to retain staff during the pandemic. The student wrote a title and headings in his own words and pasted the hardships in bullet point form from the other chatbot to the composition (turn 2). Although 18 minutes in length, we note that the student made two edits, and did not produce any recognizable expository writing units in the screen recording.

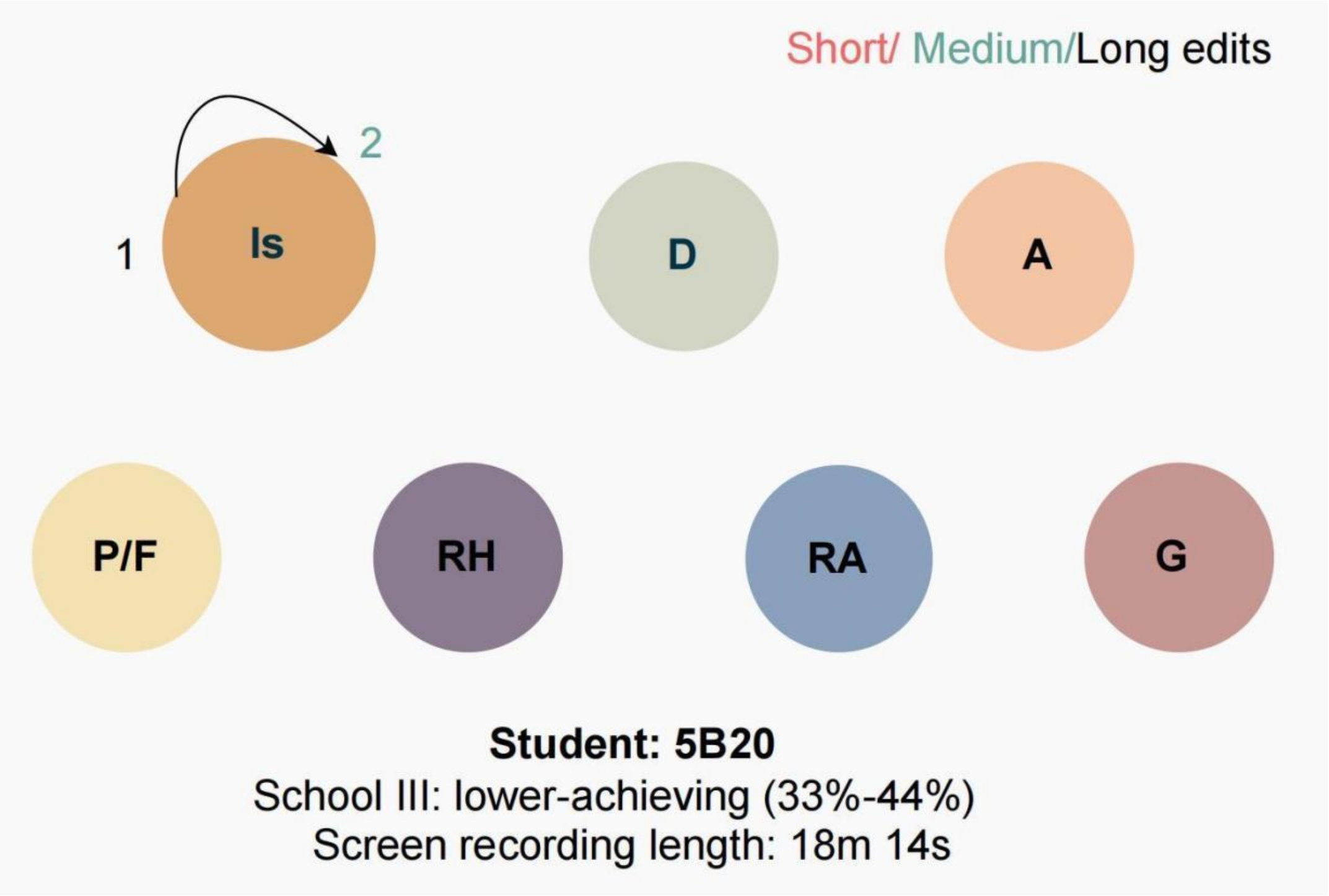

**Figure 15.** A third case of planning with bottom-up drafting and revising

*4.2.4. Bottom-up drafting and revising without planning (n=3)*

Figure 16 illustrates the case of School II student 4B10's editing. This case features *a bottom-up drafting and revising* theme shared with the cases in Figures 13-15, but *without explicit planning*. The screen recording begins with the student writing the article title in her own words, and then writing the introductory paragraph with her own words, short chunks pasted from AI chatbot output (turns 1, 3) and Google's next word prediction (turn 2). The student composed the topic sentence of the first body paragraph using her own words and a short chunk pasted from AI chatbot output (turn 4) and did the same to compose the first body paragraph's supporting sentences (turn 5). The student then composed the second body paragraph topic sentence (turns 6, 7) and supporting sentences (turns 8-11) with her own words, a short chunk from AI chatbot output and Google's next word prediction. The student next composed the third body paragraph topic sentence (turn 12) and supporting sentences (turns 13, 14), followed by a fourth body paragraph (turns 15-18).

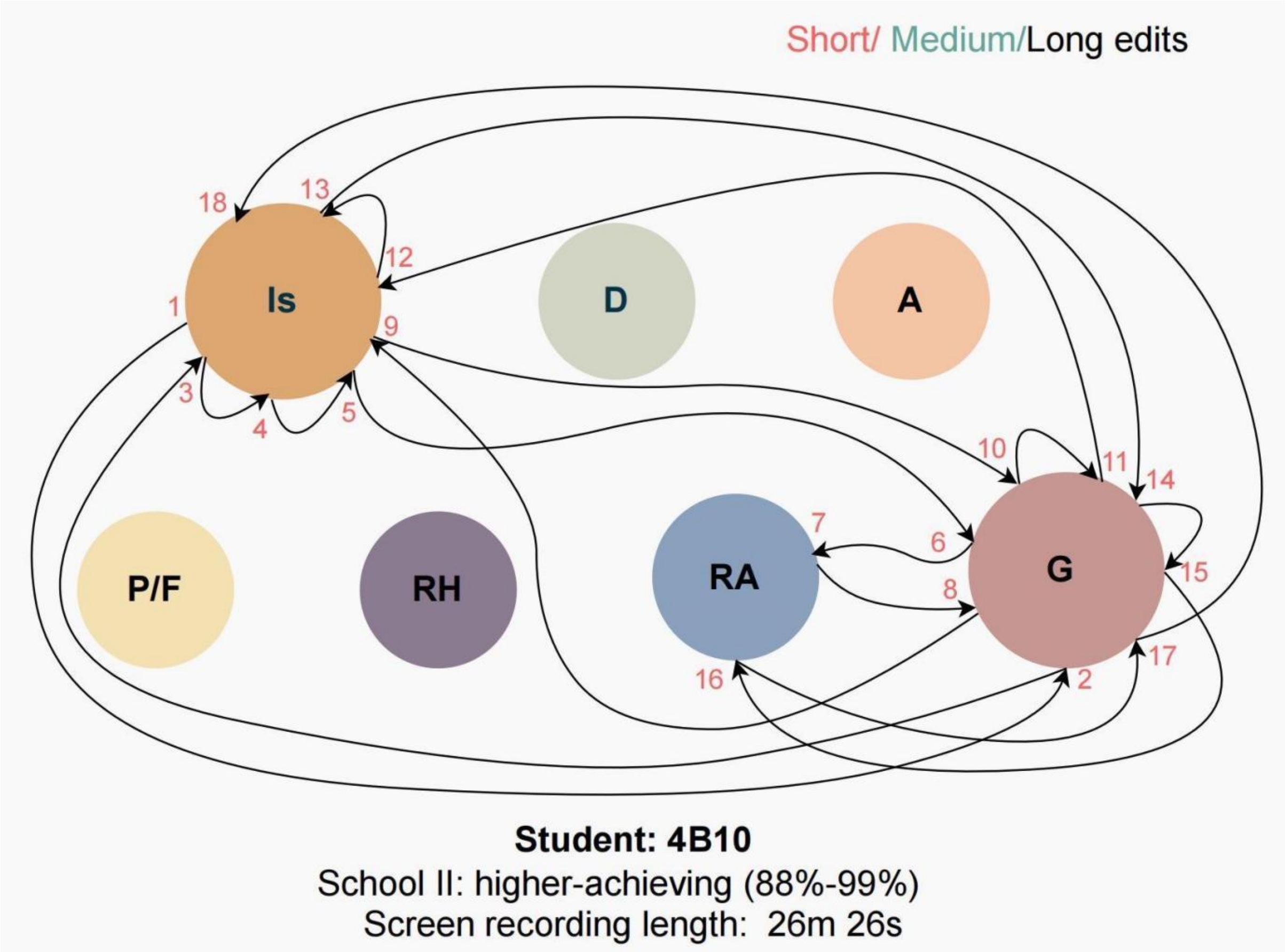

**Figure 16.** A case of bottom-up drafting and revising without planning

## 5. Discussion

### *5.1. Major Findings*

#### *5.1.1. Types of students' AI-generated text editing*

We identified 15 types of AI-generated text insertions, deletions and modifications across seven categories. Our taxonomy reveals nuanced and sophisticated interaction between students and AI-generated text beyond mere insertion of such text in a composition. Besides, the high frequency of AI-generated text insertions, deletions and replacements indicates that students are actively evaluating AI-generated text, not passively accepting it. This challenges concerns that students merely submit unedited AI-generated compositions (Barrot, 2023).

Importantly, our findings revealed several types of AI-generated text edits initiated by iOS and Google Docs software. Specifically, EFL students' intention to use their own words with their misspellings, grammatical errors and partially completed words unintentionally invited Apple's and Google's AI-initiated text edits into a composition. These unanticipated, semi-autonomous sources of AI-generated text suggest that as generative AI becomes more autonomous (Xi et al., 2025) and ubiquitous in digital writing tools, it introduces additional complexity to students' writing process. Furthermore, generative AI's unsolicited contributions to student writing go beyond its providing solicited, timely assistance for process writing (Guo & Li, 2024; Su et al., 2023). It potentially threatens students maintaining full agency over the machine-in-the-loop writing process.

#### *5.1.2. Patterns of students' AI-generated text editing*

We identified four patterns by which students approach AI-generated text editing: planning with top-down drafting and revising, top-down drafting and revising without planning, planning with bottom-up drafting and revising and bottom-up drafting and revising without planning. The patterns represent and extend configurations of cognitive writing processes (Flower & Hayes, 1981) to machine-in-the-loop writing. For example, we found that planning with a machine-in-the-loop can involve a student prompting generative AI chatbots for information about a specific topic or requesting a writing plan before drafting. Importantly, although we found two drafting and revising approaches, the top-down approach whereby students insert a complete AI-generated composition and modify that is a major shift from traditional drafting and revising. Furthermore, the top-down approach without explicit planning shows some students are immediately evaluating and revising a complete text from the outset, rather than generating ideas and gradually developing a complete text. Nonetheless, that the majority of students adopted planning with bottom-up drafting and revising indicates many students maintain a traditional cognitive writing process and may prefer to preserve their agency when writing with a machine-in-the-loop. Our multiple case study and network graph visualizations indicate that broad AI-generated text editing patterns can show some variation within a pattern.

### *5.2. Implications*

Our findings contribute to theoretical understanding of EFL students' machine-in-the-loop writing. Our taxonomy of 15 edit types across seven categories details how students integrate AI-generated output into their composition and develop that AI-generated text. Those and our four AI-generated text editing patterns demonstrate that machine-in-the-loop writing can involve even more complex cognitive processes than traditional writing. For example, students must navigate the evaluation and modification of AI-generated text, and have the choice of doing so at the outset of drafting a composition. Besides, they must manage increasingly unsolicited AI-initiated interventions in their writing. In sum, our study's findings on EFL students' modifications and management of AI-generated text show a shift in cognitive writing processes from how students traditionally plan, revise and draft their own ideas.

Practically, our findings support Yeh's (2024) advocacy for integrating AI into EFL instruction by informing the explicit instruction of AI-generated text editing strategies in writing pedagogy. First, rather than focus on whether EFL students should use AI tools, educators can acknowledge that students can invest much time and cognitive effort to develop a composition with AI-generated text. When delivering instruction on writing with a machine-in-the-loop, educators can teach explicit planning approaches with prompt engineering skills, as well as model bottom-up and top-down drafting and revising approaches. In doing so, educators can also instruct on how to maintain authorial voice and agency in the machine-in-the-loop writing process. Beyond assessment of final compositions, educators may consider formative assessment of students' strategic and effective engagement with AI-generated text during the writing process.

### *5.3 Limitations and Future Research*

This study's Hong Kong EFL secondary school context may limit generalization to similar language learning contexts. Besides, the screen recordings were limited to one instance of each student's writing process, and did not capture the entirety of that process, such as what students did after the workshop but before submitting the composition. Therefore, the short duration of screen recordings limits our observations into how editing patterns might evolve. Future studies can investigate whether similar editing patterns emerge across diverse educational contexts, and how patterns evolve alongside advances in generative AI. Future studies can employ more diverse methods, including longitudinal studies, and think-aloud protocols to collect data on students' decision-making when writing with a machine-in-the-loop.

## 6. Conclusion

This study investigated EFL secondary school students' editing of AI-generated text to complete an expository writing task. It identified 15 types of AI-generated text edits across seven categories, and four distinct editing patterns. Our findings show EFL students' process of writing with a machine-in-the-loop can be intensive and complex, their editing going beyond the mere insertion of a complete AI-generated composition. The findings challenge assumptions about students' passive consumption of AI-generated output. They also portend increasing complications to student agency because of autonomous generative AI in the writing process. They can inform development of explicit instructional approaches to writing with AI-generated text. Ultimately, EFL educators must adapt not only their tools for the digital writing age, but also pedagogies and expectations so that students are not only equipped to be proficient writers independent of AI support, but also proficient writers with a machine-in-the-loop.